\documentclass[aps,prl,amsmath,amssymb,twocolumn,superscriptaddress,letterpaper]{revtex4}
\usepackage{dcolumn}
\usepackage{graphicx}
\usepackage{epstopdf}
\usepackage{verbatim}
\usepackage{amssymb}
\usepackage{amsmath}
\usepackage{color}

\begin{document}

\title{Theoretical Description of a DNA-Linked Nanoparticle Self-Assembly}

\author{Chia Wei Hsu}
\affiliation{Department of Physics, Wesleyan University, Middletown, Connecticut 06459, USA}

\author{Francesco Sciortino}
\affiliation{Dipartimento di Fisica and CNR-ISC, Universit\`a di Roma La Sapienza, Piazzale Aldo Moro 2, I-00185 Rome, Italy}

\author{Francis W. Starr}
\affiliation{Department of Physics, Wesleyan University, Middletown, Connecticut 06459, USA}

\date{Submitted 3 May 2010; Resubmitted 24 June 2010}

\begin{abstract}

  Nanoparticles tethered with DNA strands are promising building blocks
  for bottom-up nanotechnology, and a theoretical understanding is
  important for future development.  Here we build on approaches
  developed in polymer physics to provide theoretical descriptions for
  the equilibrium clustering and dynamics, as well as the self-assembly
  kinetics of DNA-linked nanoparticles.  Striking agreement is observed
  between the theory and molecular modeling of DNA tethered
  nanoparticles.
\end{abstract}

\maketitle


The specificity, directionality, and technological control over base
sequence make DNA an attractive linking unit for artificial
constructs~\cite{Niemeyer00,Seeman03,Condon06}.  One such approach is to
attach DNA strands of designed base sequence onto a nanoparticle (NP),
thereby creating ``molecules'' that self-assemble into highly organized
structures through complementary pairings of
DNA~\cite{Mirkin96,Alivisatos96}.  Recent experiments have demonstrated
that uniformly DNA-functionalized NP can self-assemble into
disordered~\cite{crocker03,hk05,gang07,eiser} or ordered crystal
structures~\cite{Mirkin08,Gang,Crocker09,Mirkin09}.
NPs with a discrete number of attachments are harder to prepare, but
they provide possibilities for even more diverse
structures~\cite{Stewart, Alivisatos2001,Suzuki09,hlss}.  In addition to
examining the possible equilibrium structures, it is also vital to
develop an understanding for the kinetics of the self-assembly
process~\cite{gang07,Crocker09,Mirkin09,Crocker05-PRL,Dreyfus09}, since
the pathway to desired structures may depend sensitively on sample
preparation.

In this Letter, we build on concepts from polymer physics to provide a
theoretical framework that describes both the equilibrium properties and
the self-assembly kinetics of NPs functionalized with a small number of
ssDNA.  We demonstrate the applicability of this theory to molecular
simulations of a binary mixture where NPs are functionalized with two or
three ssDNA, similar to experimentally realized
systems~\cite{Alivisatos2001}.

\begin{figure}[th]
\centerline{
\includegraphics[width=3.4in]{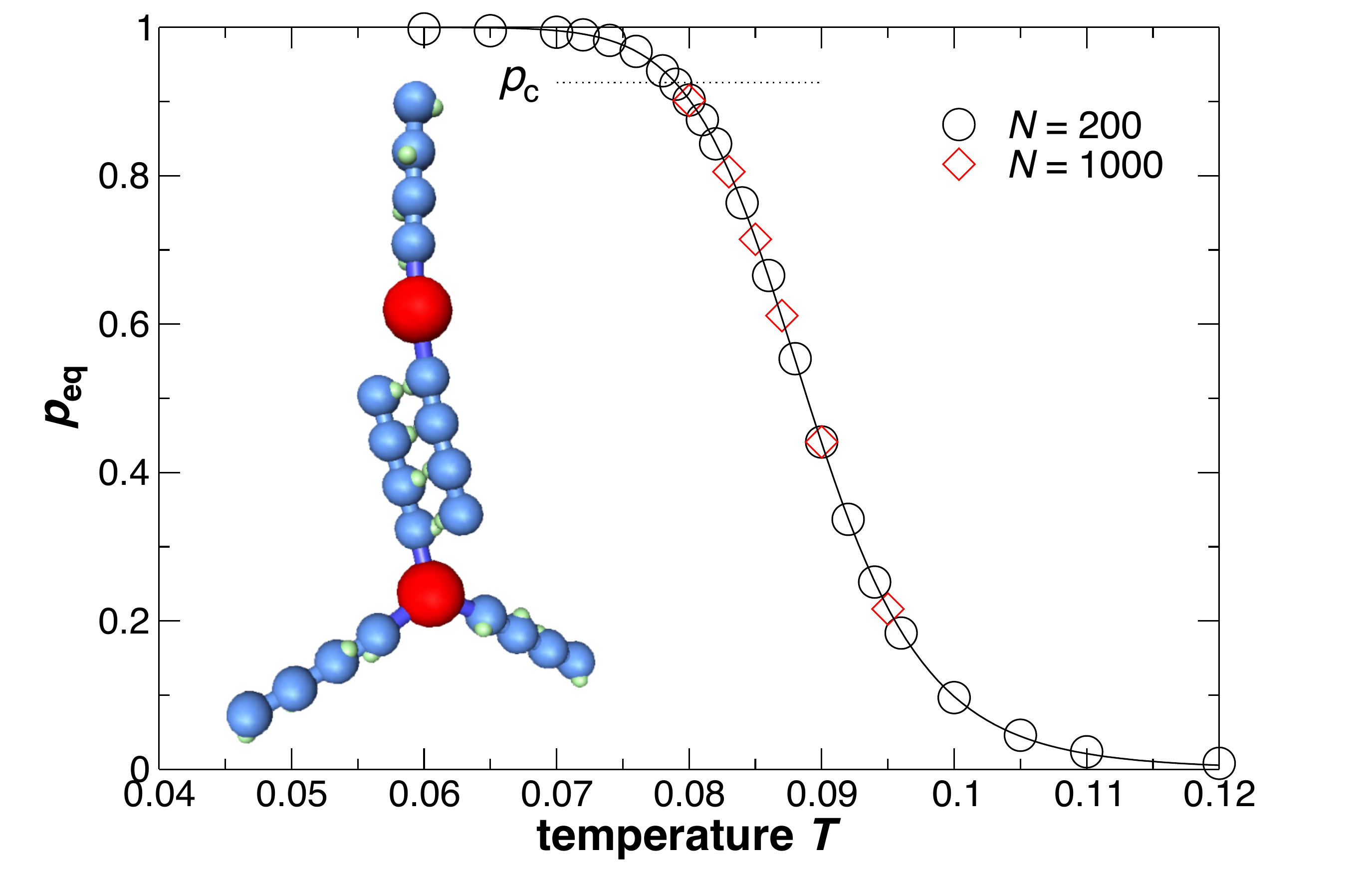}
}
\vspace{-12pt}
\caption{Temperature dependence of the fraction of bonded arms at
  equilibrium, $p_{\mathrm{eq}}$. Symbols are simulation data for system
  size $N=200$ and $N=1000$. Solid line represents two-state expression
  with $\Delta U=1.78$ and $\Delta S=20.0$. Dotted line indicates
  percolation threshold predicted by FS theory. Inset shows graphical
  representation of one 3-functional particle bonding to one
  2-functional particle. Large red spheres indicate NP, blue spheres
  indicate backbone beads of ssDNA, and small green spheres indicate
  location of ``sticky'' sites.}
\label{fig:pb-T}
\end{figure}

The coarse-grained molecular model we use for the DNA-functionalized NP
is a modest modification of a previous model~\cite{ss}.  The inset of
fig.~\ref{fig:pb-T} provides a graphical representation. The
sugar-phosphate backbone is represented by connected beads with only
excluded volume interactions; each bead carries an additional ``sticky''
site that represents a base (A, T, C, or G) and can bond only with
another sticky site of complementary type (A-T or C-G pair).  The size
of the sticky site is small relative to the backbone beads, so that a
base can bond to at most one other complementary base. A bending
potential between consecutive triplets of backbone beads is included to
model the characteristic rigidity of the DNA strands and to maintain the
angle between ssDNA attached to the same core ($120\,^{\circ}$ for
3-functional NP and $180\,^{\circ}$ for 2-functional NP).  A more
complete description of the model is provided in Ref.~\cite{ss}.  We
consider ssDNA four bases in length with sequence A-C-G-T, chosen to
enable complementary pairing between ssDNA on different NPs. To mimic
solvent effects on dynamics, we simulate this model using dissipative
particle dynamics~\cite{Groot_Warren}, an algorithm known to ensure
correct hydrodynamic behavior.  Length is measured in diameter of the
beads, and temperature $T$ is measured in units of the bonding energy
between complementary bases, taking Boltzmann's constant $k_B=1$ for
simplicity.

We consider a system with $N_2=945$ 2-functional particles and $N_3=55$
3-functional particles.  The presence of 3-functionalized particles
facilitates branching of otherwise linear chains, thereby opening the
possibility to form a percolating network and phase separation at low
density~\cite{bianchi}. For part of our study where system size is less
important, we consider a smaller system with $N_2=189$ and $N_3=11$ so
that we may more readily study the dynamics of the most slowly relaxing
systems.  Nonetheless, equilibration at the lowest $T$ studied in the
small system still requires $5 \times 10^9$ integration steps -- roughly
3 months of computation on a single core with current resources.  The
number density is fixed at $(N_2+N_3)/V=0.0065$ (corresponding to mean
separation of 5.4 bases between particles); we have confirmed with
structural factor $S(q)$ that there is no (gas-liquid) phase separation
at this density.


In this system, the extent of reaction $p$ is characterized by the
fraction of bonded strands.  We define two strands as bonded when half
or more of the base pairs on them are linked ({\it i.e.} the base-base
potential energy is negative).  The $T$ dependence of equilibrium $p$ is
well described (see fig.~\ref{fig:pb-T} and Ref.~\cite{ss}) by a
two-state expression
\begin{equation}
p_{\mathrm{eq}}(T)=\left[1+\exp\left({-\frac{\Delta U - T \Delta S}{k_B T}}\right)\right]^{-1},
\label{eq:pb-T}
\end{equation}
where $\Delta U$ and $\Delta S$ measure, respectively, the energy and
entropy change associated with the formation of a double strand.

\begin{figure}[th]
\begin{center}
\includegraphics[width=3.4in]{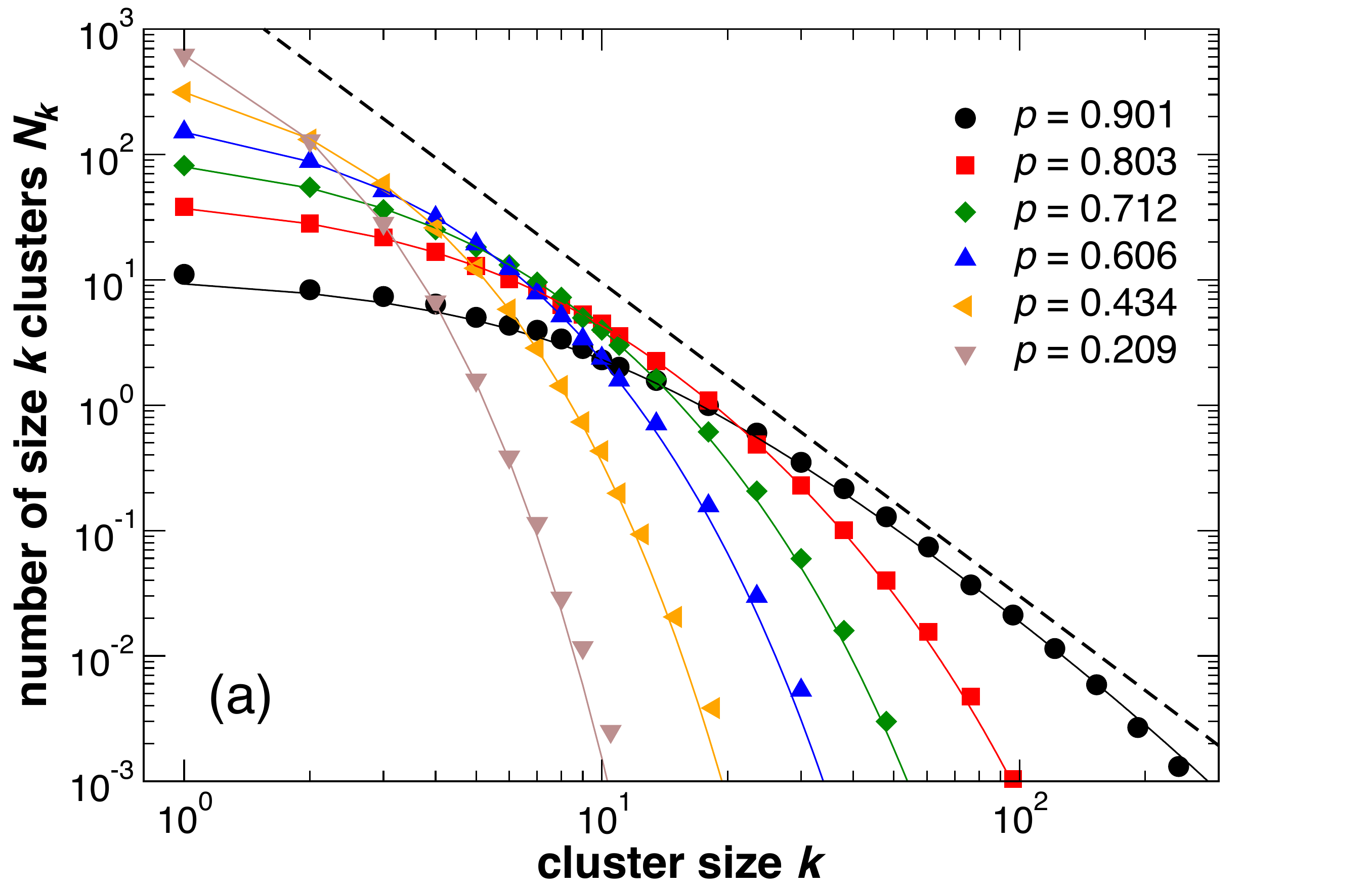}
\includegraphics[width=3.4in]{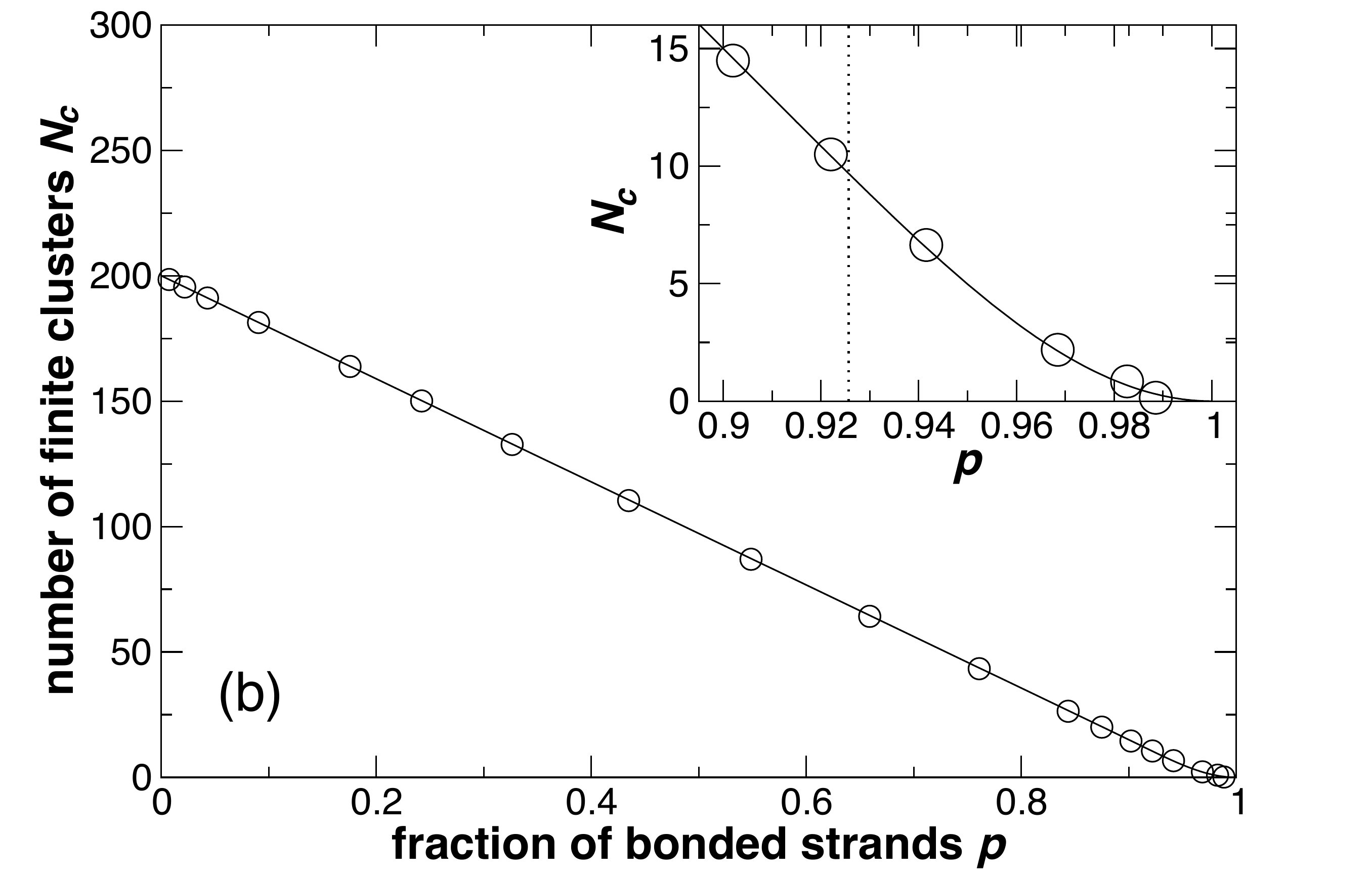}
\end{center}
\vspace{-15pt}
\caption{(a) Equilibrium distribution of finite-size clusters
  $N_k$. Points are simulation data (system size $N=1000$) and lines are
  the FS predictions. Dashed line represents the expected power-law
  decay $N_k \sim k^{-2.5}$ at percolation~\cite{RubinsteinBook}. (b)
  Number of finite size clusters $N_c$ as a function of $p$. Points are
  simulation data (system size $N=200$) and black solid line is FS
  prediction.  Inset enlarges the region above percolation threshold
  (indicated by dotted line).}
\label{fig:csd}
\end{figure}

To quantify the equilibrium assembly, we first calculate the number of
clusters $N_k$ of size $k$ at each $T$ (fig.~\ref{fig:csd}(a)).  To
describe our results, we consider the theoretical predictions of
Flory~\cite{flory-book} and Stockmayer~\cite{stockmayer1943} (FS). The
FS theory assumes that no intra-cluster loops are present and that every
functional group is equally reactive.  The theory predicts the most
probable number of clusters $N_{nl}$ containing $n$ 3-functional
particles and $l$ 2-functional particles is
\begin{equation}
\begin{split}
N_{nl}(p) &= N_3 (1-p)^{n+2} (p_3p)^{n-1} (p_2p)^{l} w_{nl},\\
w_{nl} &= 3 \frac{(2n+l)!}{n!l!(n+2)!},
\end{split}
\label{eq:csd}
\end{equation}
where $p_3=3N_3/(2N_2+3N_3)$ and $p_2=1-p_3$ are the portions of
functional groups belonging to 3-functional or 2-functional particles,
respectively.  A sum yields the analytic prediction
$N_k=\sum_{n+l=k}N_{nl}$.  Since the only variable in eq.~\ref{eq:csd}
is the extent of reaction $p$ (which we already know), we have a
parameter free description of $N_k$.  We find quantitative agreement
between this prediction and our results (fig.~\ref{fig:csd}(a)).  We can
also evaluate the number of finite-size clusters $N_c=\sum_{k}N_k$,
which we compare with the corresponding simulation results in
fig.~\ref{fig:csd}(b).  We again find quantitative agreement in the
whole range of $p$ -- both below and above the percolation threshold
$p_c = 1/(1+p_3)=0.9256$ -- a demonstration that in this system an
accurate analytic description for the equilibrium clustering is
possible.  Qualitatively, these clusters are dominated by long chains of
2-functional NP, with occasional links provided by the 3-functional NP.

\begin{figure}[th]
\begin{center}
\includegraphics[width=3.4in]{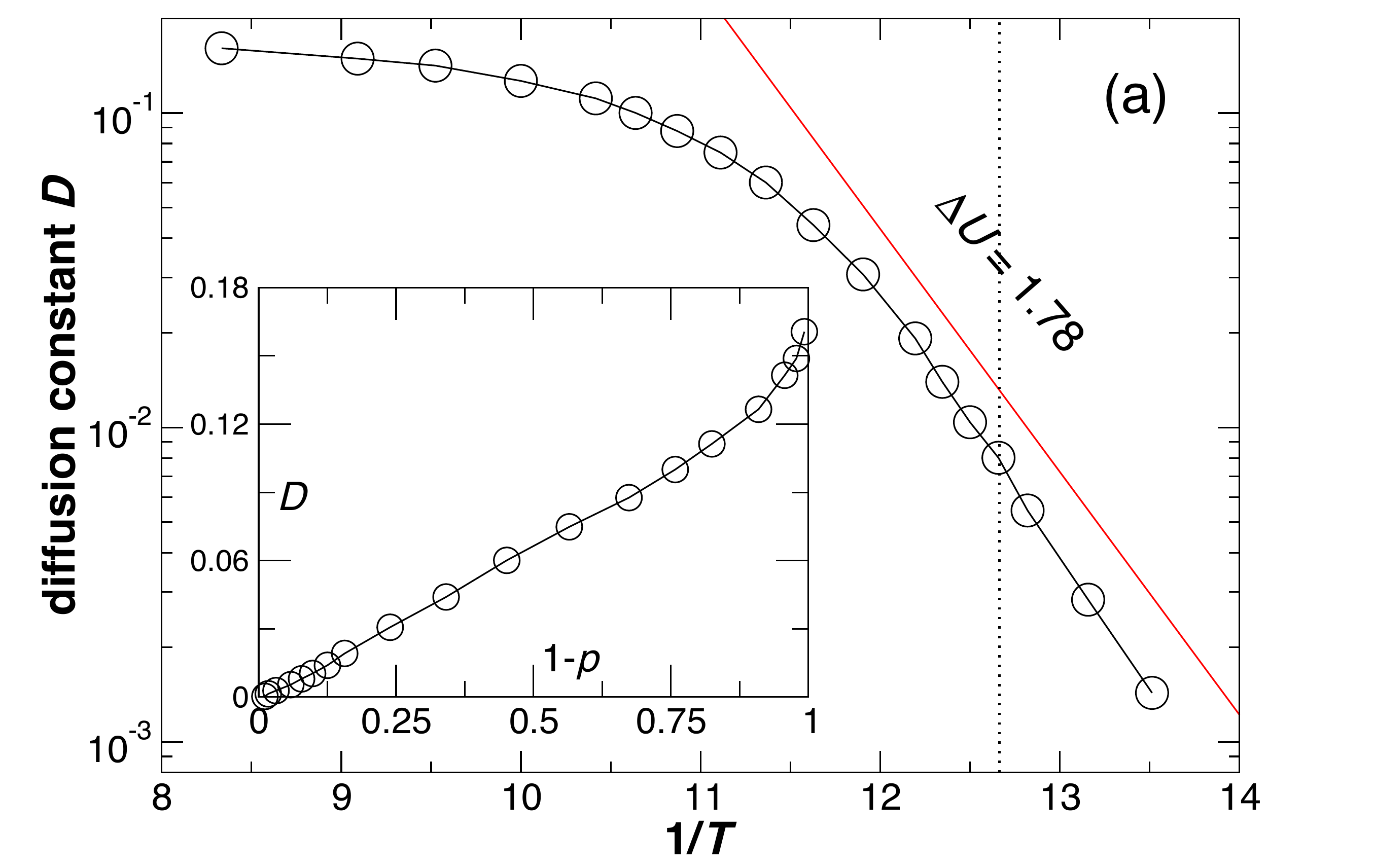}
\includegraphics[width=3.4in]{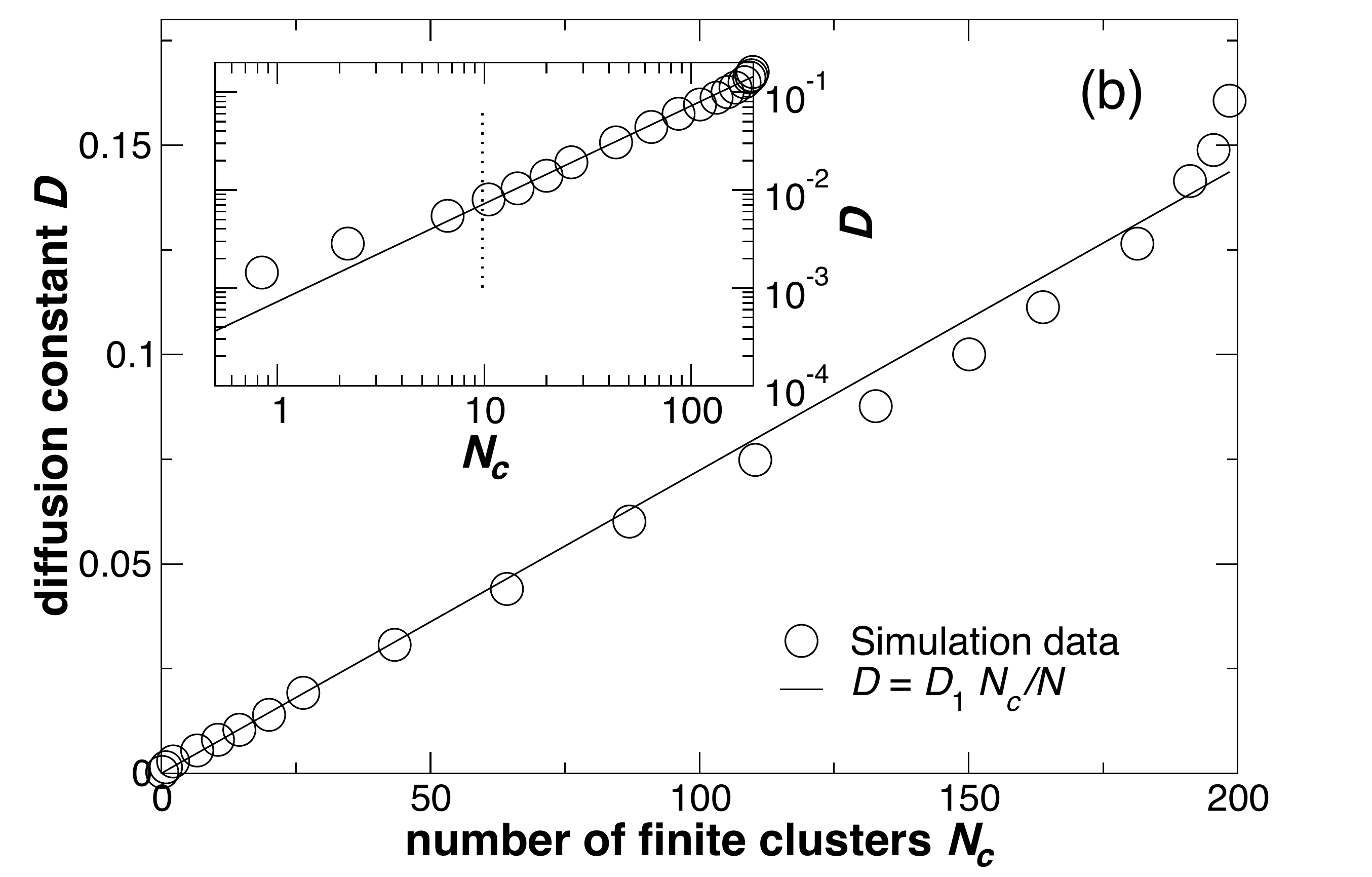}
\end{center}
\vspace{-15pt}
\caption{(a) Diffusion constant $D$ as a function of inverse
  temperature. The dotted line indicates the location of the percolation
  threshold, and red solid line represents Arrhenius behavior with an
  activation energy $\Delta U=1.78 $ -- the same value reported for
  $p(T)$ in fig.~\ref{fig:pb-T}.  See text for an explanation.  Then
  inset shows $D$ as a function of fraction of non-bonded arms.  (b)
  Diffusion constant $D$ versus number of finite clusters $N_c$.  The
  line shows the predicted behavior $D = D_1\; N_c/N$, where $D_1 =
  0.145$. Calculation of $\langle \Delta r^2 \rangle$ is averaged over
  $8$ to $15$ independent runs. System size $N=200$.}
\label{fig:D-Nc}
\end{figure}

The clustering process has a profound effect on the dynamics of the
system, as revealed by the diffusion constant $D$, which drops several
orders of magnitude in the small $T$-region where clusters form.
Figure~\ref{fig:D-Nc}(a) shows $D$ evaluated from the asymptotic
behavior of the mean-square displacement $\langle \Delta r^2 \rangle =
6Dt$ as a function of the inverse temperature $1/T$.  Close to the
percolation transition, $D$ takes on an Arrhenius $T$ dependence.

To understand the behavior of $D$, consider that the overall diffusion
constant is an average over particles belonging to clusters of different
sizes; therefore $D=\sum_k D_k (k N_k)/N$, where $D_k$ is the diffusion
constant for particles in a cluster of size
$k$~\cite{conigliojcp2009}. Typically, $D_k$ drops inversely to the
cluster size, so that $D_k \approx D_1/k$, the so-called Stokesian limit
of diffusion~\cite{Oshanin_Moreau,Guzman}. In this approximation, $D
\approx {D_1} \sum_k N_k/N = D_1 N_c/N$.  If we take the monomer
diffusion coefficient $D_1$ as a constant, we can readily test this
prediction. Figure~\ref{fig:D-Nc}(b) shows that $D$ obeys this
prediction well, with small deviations at the largest and smallest
values.  Numerically, we find $D_1 = 0.145$, roughly equal to the high
$T$ asymptotic value of $D$, where monomers dominate. As shown in
fig.~\ref{fig:csd}(b), $N_c \sim (1-p)$ for $p<p_c$, and so we also
expect $D \sim (1-p)$ to leading order (fig.~\ref{fig:D-Nc}(a) inset).
Since at low $T$, $(1-p) \sim \exp(-\Delta U/k_B T)$ (from
eq.~\ref{eq:pb-T}), we can explain the Arrhenius behavior $D \sim
\exp(-\Delta U/k_B T)$ in the low $T$ region.  Indeed, $\Delta U$
obtained from $p(T)$ (fig.~\ref{fig:pb-T}) matches that for $D$,
demonstrating consistency.

Having successfully described the clustering and dynamics at
equilibrium, we next explore to what degree the self-assembly kinetics
can be analytically predicted.  We focus on the evolution of the system
after a temperature jump (at $t=0$) from a high-$T$ unassociated state
to a low-$T$ self-assembled state, following the evolution of $p$ from
its initial value $p(t=0) \approx 0$ to its final equilibrium value
$p(t=\infty)=p_{eq}(T)$.  With the assumption that every strand is
equally reactive independent of the size of cluster to which it
attaches, $p(t)$ satisfies~\cite{dongen}
\begin{equation}
\frac{dp}{dt} =   p_{\mathrm{eq}} k_{\mathrm{overall}} \left[ \frac{\left(1-p\right)^2}{\left(1-p_{\mathrm{eq}}\right)^2}  - \frac{p}{p_{\mathrm{eq}}} \right],
\label{eq:dpdt}
\end{equation}
where $k_{\mathrm{overall}}$ is the overall rate coefficient for bond
breaking. Equation~\ref{eq:dpdt} simply states that the rate of bond
formation (first term) only depends on the probability of finding two
unassociated strands, and that the rate of bond fragmentation (second
term) only depends on the probability of finding a bonded strand.  When
this kinetic process is dominated by the intrinsic rate of reaction
({\it i.e.} dsDNA formation), $k_{\mathrm{overall}}=k_c$ is a constant,
and eq.~\ref{eq:dpdt} can be integrated analytically to give the time
evolution
\begin{equation}
p(t)=p_{\mathrm{eq}}\frac{1- \Lambda \mathrm{e}^{-{\Gamma}t}}{1 - p_{\mathrm{eq}}^2 \Lambda  \mathrm{e}^{-{\Gamma}t}},
\label{eq:pt-original}
\end{equation}
where $\Gamma=k_c (1+p_{\mathrm{eq}})/(1-p_{\mathrm{eq}})$ determines
the rate of reaction, and $\Lambda =
(1-p(0)/p_{\mathrm{eq}})/(1-p(0)p_{\mathrm{eq}})$ incorporates the
initial condition $p(0)$.  It has been shown by van Dongen and
Ernst~\cite{dongen} that, in this so-called ``reaction-limited'' or
``chemical-limited'' process, eq.~\ref{eq:dpdt} provides a time
dependent distribution $N_k(p(t))$ (together with eq.~\ref{eq:csd}) that
satisfies the Smoluchowski rate equation with condensation and
fragmentation terms below percolation~\cite{francesco-jpc10,footnote}.

\begin{figure}[th]
\begin{center}
\includegraphics[width=3.4in]{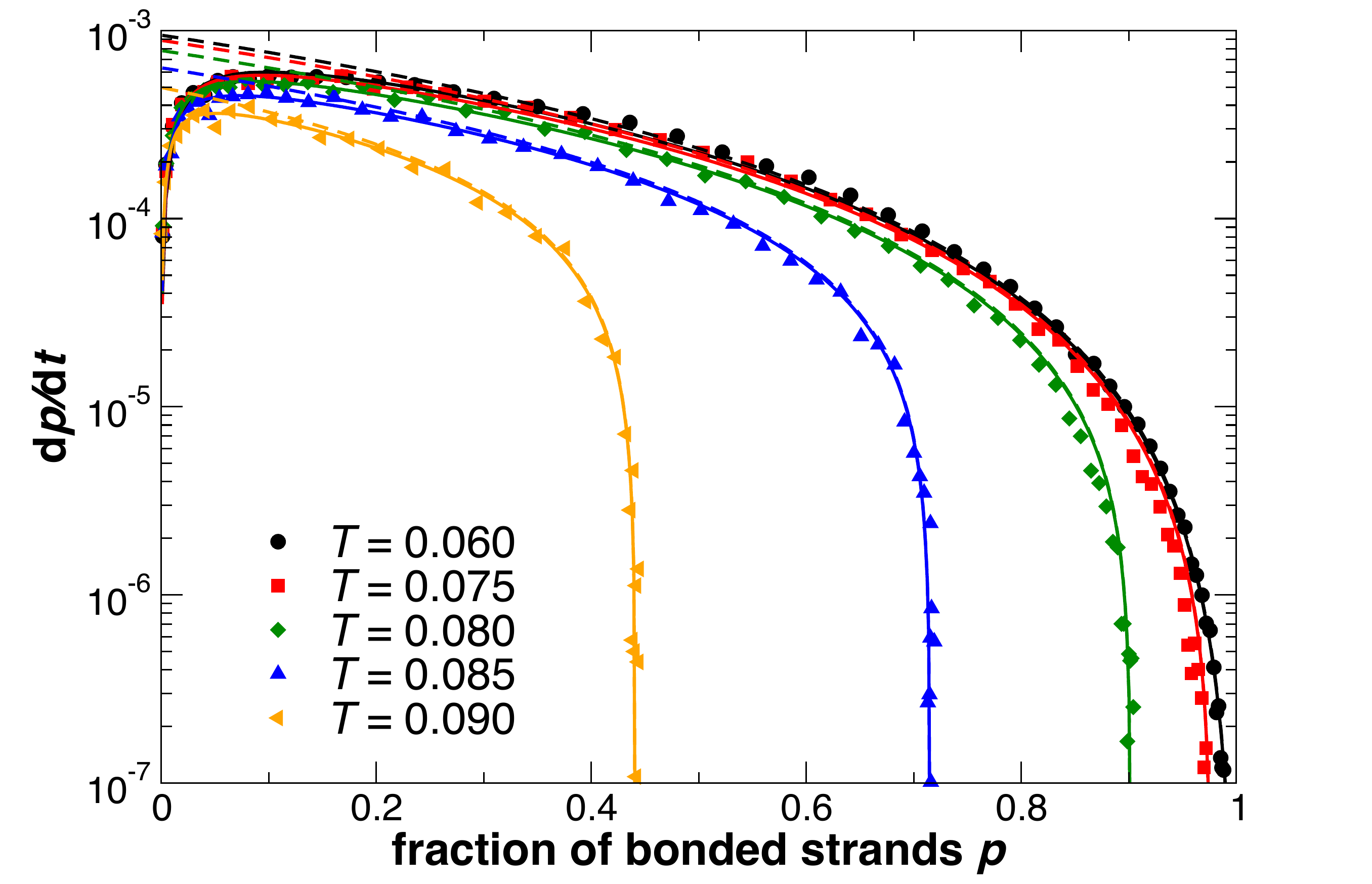}
\includegraphics[width=3.4in]{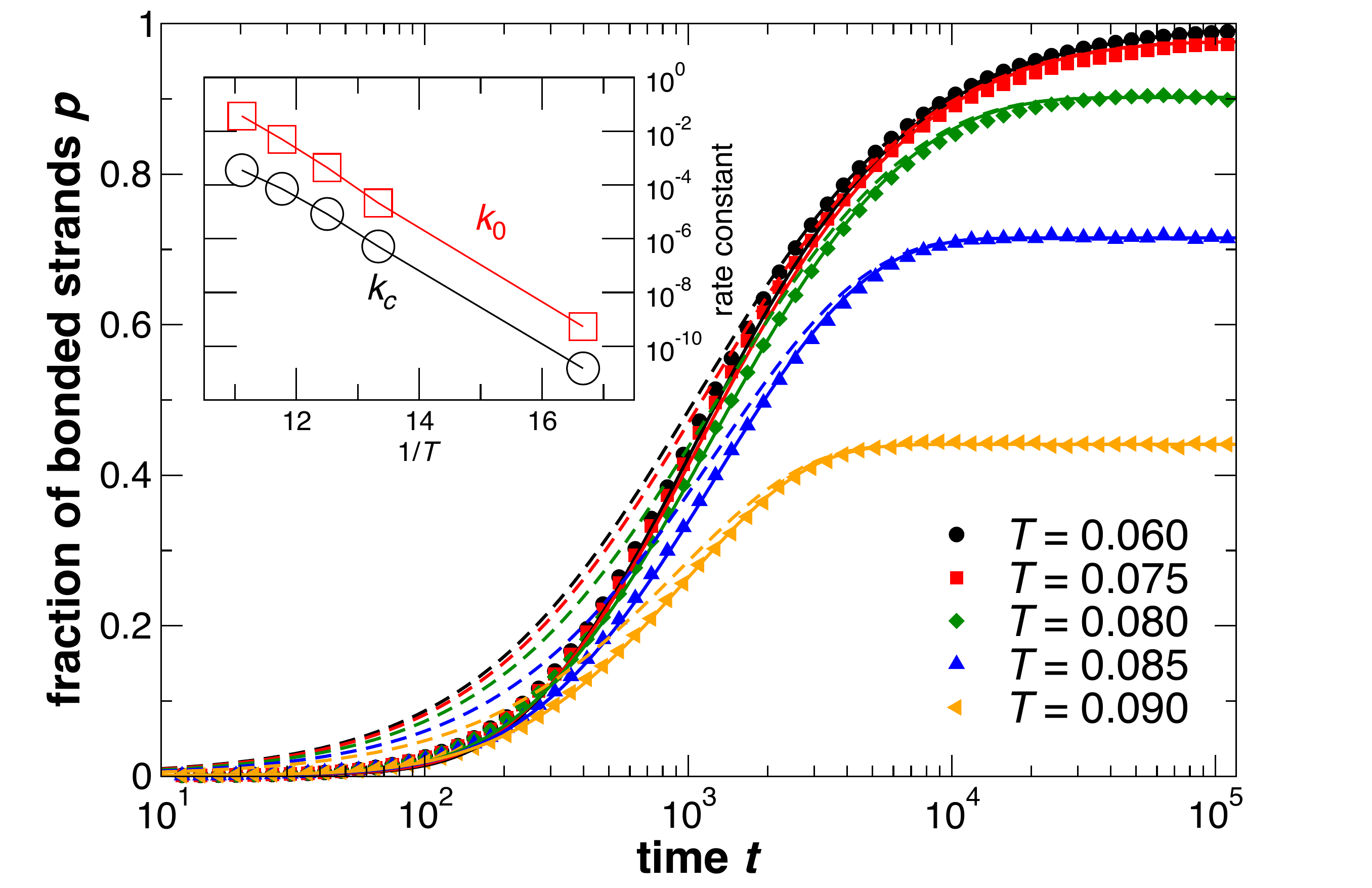}
\end{center}
\vspace{-15pt}
\caption{(a) Rate of bond $dp/dt$ as a function of fraction $p$ of
  bonded strands, and (b) time evolution of $p$ starting from a high-$T$
  state ($T_{\mathrm{init}}=0.2$, $p(0)=0.008$). Inset shows the two
  rate constants from best fit. Symbols are simulation results (system
  size $N=1000$), averaged over $35$ independent runs at short time and
  $15$ runs at longer time.  Dashed and solid lines represent,
  respectively, prediction of eq.~\ref{eq:dpdt} in the reaction-limit
  scenario (constant $k$) and with consideration of the $p k_0$ rate
  coefficient.}
\label{fig:pb-t}
\end{figure}

Using our simulation results, we test if our system can be described by
the reaction-limited description given by eq.~\ref{eq:dpdt} with
$k_{\mathrm{overall}}=k_c$.  In fig.~\ref{fig:pb-t}, we see that the
reaction-limited expression describes the rate of bond $dp/dt$ and time
evolution $p(t)$ well for most values of $p$ but fails when $p \lesssim
0.3$.  This indicates that, in the assembly process of
DNA-functionalized NP, there is a rate-controlling factor for small $p$
not captured by a simple reaction-limited description. The trend in
$dp/dt$ shows that this additional rate-controlling factor slows down
the overall reaction rate at small $p$.  To incorporate the effect, we
introduce an additional reaction rate $k_{\mathrm{small}}=k_0 p$ that
changes linearly with $p$. Since reaction rates sum
inversely~\cite{Oshanin_Moreau},
$k_{\mathrm{overall}}^{-1}=k_c^{-1}+(k_0 p)^{-1}$.
Figure.~\ref{fig:pb-t} shows that this modified $k_{\mathrm{overall}}$
accurately describes our simulation results for the assembly kinetics
over the entire self-assembly process. By numerically integrating
eq.~\ref{eq:dpdt} using the appropriate initial condition $p(0)$, we
also provide an accurate estimate of the time evolution $p(t)$ based on
this expression. Surprisingly, this prediction works well even for the
case in which the final state percolates ($T=0.06$). The inset in
fig.~\ref{fig:pb-t} shows that both rate constants $k_c$ and $k_0$
follow an Arrhenius behavior ({\it i.e.} $\ln k \sim {1/T}$).

The presence of a second rate constant that is dominant at small $p$ may
be a consequence of the fact that bonding NP via dsDNA requires two
steps: (i) the strands of opposing NP must be properly oriented, and
(ii) once oriented, the linking of individual bases must occur.  Since
base-pair bonds are dynamic, there is a constant ``flickering'' of
bonds.  Thus, when $p$ is significant (here, apparently $p \gtrsim
0.3$), the flickering should be a consequence of the regular breaking
and reforming of individual base-pair bonds.  As a result, the strands
will typically remain oriented to facilitate base-pair bonding, making
the rate constant associated with orientation irrelevant ({\it i.e.}
$k_{\mathrm{overall}}$ dominated by $k_c$).  For very small $p$, the
intermittency of the bonds is large enough that strands can ``wander'',
and as a result the time needed to orient two strands becomes the rate
limiting factor, and so $k_{\mathrm{small}}$ dominates.  This is
consistent with the fact that for simpler systems, where arm orientation
is not a factor, there is only a single rate constant to
consider~\cite{francesco-softmatter09}. The effect of this mechanism is
the opposite of the diffusion-limited scenario, which is usually
important only for long-time dynamics, when $p$ is
large~\cite{Oshanin_Moreau,francesco-jpc10}.

Our studies show that, by combining a thermodynamic description for
$p_{\mathrm{eq}}(T)$, the FS theory for $N_k(p)$, an approximate
expression for $D$, and rate equation with appropriate rate constants,
we can provide a comprehensive theoretical description for the
clustering, equilibrium dynamics, and assembly kinetics that accurately
predicts the behavior of this DNA-functionalized NP system. The complex
process of self-assembly through dsDNA formation can be simplified as a
chemical-limited reaction plus an additional reaction rate $k_0 p$
associated with strand orientation.
An obvious question is whether the results obtained for our simplified
DNA simulation model will be transferable to real systems.  The
experimentally synthesized systems with very small numbers of attached
DNA strands to a core NP studied by Alivisatos and
co-workers~\cite{Alivisatos2001} offer the possibility to directly test
our predictions.  Additionally, our approach could be applicable to the
experiments of Gang and co-workers~\cite{gang07} where there are many
strands attached to a core NP, but only a small number of which are
available for binding, resulting in more elongated clusters.  They also
consider micron-sized colloidal cores in that work, which can
potentially significantly alter the assembly dynamics.

The successful theoretical description of our numerical results shows
that existing frameworks can be adapted describe novel self-assembled
systems.  The next step in this development is the satisfactory
inclusion of closed loop structures, which is important for applications
when the number of functionalizing strands is larger.

We thank J.~Douglas for discussion, and Wesleyan University for computer
time, which was supported by National Science Foundation Grant
CNS-0959856. This work was supported by National Science Foundation
Grant DMR-0427239. FS acknowledges support from
ERC-226207-PATCHYCOLLOIDS and NoE SoftComp NMP3-CT-2004-502235.


\begin{thebibliography}{10}
\bibitem{footnote} Ref.~\cite{dongen} showed the case for
  $\mathrm{RA}_{f}$ particles; the proof for a binary mixture of
  $\mathrm{RA}_{2}$ and $\mathrm{RA}_{f}$ particles can be done in a
  similar way.
\end{thebibliography}
\end{document}